# Room-temperature polariton condensate in a two-dimensional hybrid perovskite

Marti Struve[1], Christoph Bennenhei[1], Hamid Pashaei Adl[1,2], Kok Wee Song[3], Hangyong Shan[1], Nadiya Mathukhno[1], Jens-Christian Drawer[1], Falk Eilenberger[4,5], Naga Pratibha Jasti[6,7], David Cahen[7], Oleksandr Kyriienko[3], Christian Schneider[1], Martin Esmann[1,+]

[1]*Institut für Physik, Fakultät V, Carl von Ossietzky Universität Oldenburg, 26129 Oldenburg, Germany*

[2]*Instituto de Ciencia de los Materiales, University of Valencia, 46980 Valencia, Spain*

[3]*Department of Physics and Astronomy, University of Exeter, Exeter EX4 4QL, United Kingdom*

[4]*Fraunhofer-Institute for Applied Optics and Precision Engineering IOF, 07745 Jena, Germany*

[5]*Institute of Applied Physics, Abbe Center of Photonics, Friedrich Schiller University, 07745 Jena, Germany*

[6]*Department of Chemistry, Bar-Ilan Univ. Ramat Gan 5290002, Israel*

[7]*Department of Molecular Chemistry and Materials Science, Weizmann Institute of Science, Rehovot 7610001, Israel*

[+]*Corresponding author. Email: m.esmann@uni-oldenburg.de*

## ABSTRACT

Layered 2D halide perovskites are chemically synthesized realizations of quantum well stacks with giant exciton oscillator strengths, tunable emission spectra and very large exciton binding energies. While these features render 2D halide perovskites a promising platform for room-temperature polaritonics, bosonic condensation and polariton lasing in 2D perovskites have so far remained elusive at ambient conditions. Here, we demonstrate room-temperature cavity exciton-polariton condensation in mechanically exfoliated crystals of the 2D Ruddlesden-Popper iodide perovskite $(BA)_2(MA)_2Pb_3I_{10}$ in an open optical microcavity. We observe a polariton condensation threshold of $P_{th} = 6.76\,fJ$ per pulse and detect a strong non-linear response. Interferometric measurements confirm the spontaneous emergence of spatial coherence across the condensate with an associated first-order autocorrelation reaching $g^{(1)} \approx 0.6$. Our results lay the foundation for a new class of room-temperature polariton lasers based on 2D halide perovskites with great potential for hetero-integration with other van-der-Waals materials and combination with photonic crystals or waveguides.

## INTRODUCTION

Hybrid organic-inorganic halide perovskites (HaPs) have recently gained considerable attention driven by fast-paced progress in photonic, optoelectronic, and photovoltaic applications [1–4]. A particularly intriguing class of these hybrid materials are layered 2D HaPs, chemically synthesized realizations of multi-quantum well (QW) stacks, represented by inorganic layers sandwiched between organic spacers (see Fig. 1b, bottom) [5–9]. Since the spacers act as potential barriers for electronic excitations in the HaP stack, low-energy electronic excitations are well confined in the inorganic parts. 2D HaPs thus share strong structural similarities with conventional multi-QW stacks, but feature significantly larger exciton binding energies. In fact, excitonic correlations are so prominent in 2D HaPs that almost all of their optical features are defined by exciton physics even at room temperature [10–14]. Just like in multi-QW stacks, the overall exciton oscillator strength in 2D HaP crystals can be scaled up by the number of layers, which is a distinct advantage over transition metal dichalcogenides (TMDCs) where tedious mechanical assembly is required for multi-layers. In addition, choosing the thickness of the inorganic layers in terms of the number of unit cells, *n*, results in a unique control over the resulting exciton resonances [7,15–17]. Unlike their 3D counterparts, 2D HaPs can even be micromechanically cleaved and re-assembled in a well-defined manner [4,18–21].

The giant excitonic oscillator strengths and binding energies, alongside with spectral tunability and versatility for complex hetero-integration render 2D HaPs an ideal material class for exploring room temperature polaritonics. The formation of cavity exciton-polaritons based on these materials has recently been observed in planar dielectric, metallic and hybrid cavities [22–28]. However, in 2D HaPs, the bosonic condensation of exciton-polaritons [29,30] - their most significant feature for applications based on their quantum coherence - has so far only been demonstrated in $(BA)_2PbI_4$ ($n = 1$, BA=butylammonium) at 4 Kelvin [31], and remained elusive at room temperature.

Here, we demonstrate room-temperature cavity exciton-polariton condensation based on mechanically exfoliated crystals of the 2D Ruddlesden-Popper perovskite (RPP) $(BA)_2(MA)_{n-1}Pb_nI_{3n+1}$ ($n = 3$) integrated into an open optical microcavity [28,32–35]. Under strong, non-resonant optical pumping we observe a polariton condensation threshold at $P_{th} = 6.76\ fJ$ per pulse. We detect a significant blue shift of $0.36\ meV$ in the condensed phase, which we attribute to non-linear processes based on Coulomb interactions between excitons as well as phase space filling effects. Through interferometric measurements, we confirm the emergence of spatial coherence across the condensate with a value of the first-order autocorrelation reaching up to $g^{(1)} \approx 0.6$ with an associated coherence time of 1 ps. Our results lay the foundation for a new class of room-temperature polariton lasers based on 2D perovskites with great potential for hetero-integration with other van-der-Waals materials and combination with complex photonic crystals.

**RESULTS**

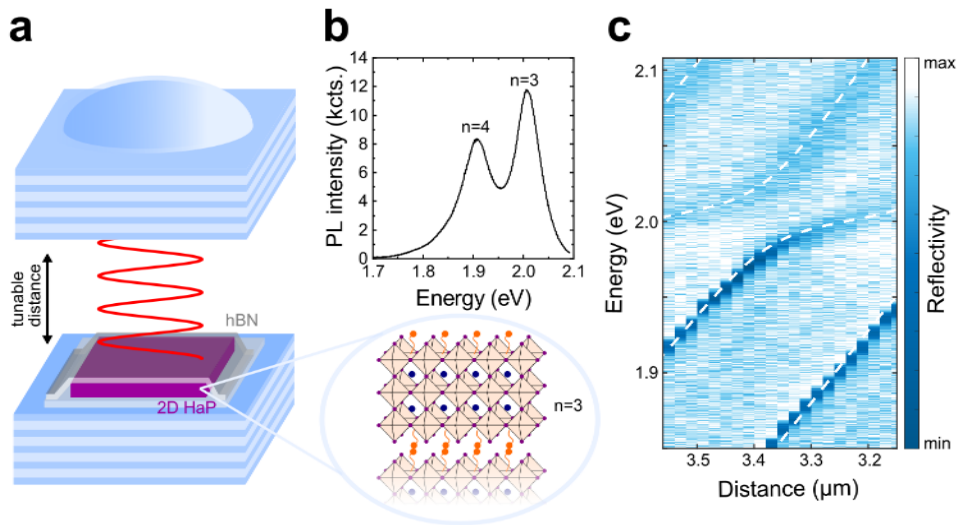

*Figure 1: **a** Two distributed Bragg reflectors (DBRs) on separate piezo stages form an open optical microcavity with tunable resonance. The upper mirror contains planar sections and sphere cap shaped indentations. The active material*

*is a mechanically exfoliated, fully hBN-encapsulated 2D halide perovskite flake. **b** (top) PL spectrum of a 218 nm thick n=3 2D HaP crystal excited at 532 nm (CW laser) containing some admixture from n=4 layers. (bottom) Structure of the 2D HaP with n=3. **c** White light (WL) reflectivity spectra as a function of the cavity air gap at normal incidence ($k_\parallel = 0$). From a coupled oscillator model (dashed lines) we deduce a light-matter coupling strength of $g = 23.5\ meV$ for the n=3 2D HaP exciton at 2eV.*

Figure 1a illustrates the implementation of our sample, based on an open optical cavity setup [32]: Two distributed Bragg reflectors (DBRs) made from alternating quarter-wave-layers of $SiO_2$ and $TiO_2$ are mounted on two separate xyz piezo stacks, forming an air-gapped Fabry-Perot resonator. In contrast to monolithic approaches, this configuration remains fully flexible in terms of cavity resonance by tuning the mirror distance and relative positioning in-plane. The top DBR contains both planar sections and sphere cap shaped indentations (6 µm diameter, 150 nm depth, cf. electron microscope image in Fig. 2a) resulting in planar Fabry-Perot and three-dimensionally confined, Laguerre-Gaussian type [36] optical modes, respectively. We mechanically exfoliated 200-300 nm thick ~50 µm x 50 µm sized flakes of chemically synthesized 2D HaPs and transferred them onto the bottom mirror by dry-gel stamping [37]. Since we operate at ambient conditions, we fully encapsulated the flakes with ~10 nm thick hexagonal Boron Nitride (hBN) layers eliminating both photo-oxidation and absorption of water [20] (see Methods section for details). The flakes that were selected for the subsequent polariton studies were n=3 2D HaPs with some admixture from n=4 layers as evidenced by the photoluminescence (PL) spectrum plotted in Fig. 1b (CW excitation at 532 nm).

To confirm strong coupling conditions in our coupled perovskite-cavity system, we performed white light (WL) reflectivity studies on our devices. In this case, a planar gold mirror (35 nm gold film thickness) was used as the top cavity reflector to optimize cavity transmission thereby enhancing the visibility of the polariton modes. Figure 1c shows the result of a WL cavity tuning series where we systematically varied the cavity air gap and recorded angle-resolved WL spectra (see Methods section). The figure shows cross-sections through these spectra extracted at $k_\parallel = 0$. We observe clearly-developed anti-crossings for subsequent longitudinal mode orders (see Supplementary Section S3 for a full range scan). At an air gap of 3.3 µm, we extract a light-matter coupling strength of $g = 23.5\ meV$ based on a coupled oscillator model (dashed lines in panel c) and based furthermore on a comparison to transfer matrix simulations with dielectric functions taken from [8] (see Supplementary Sections S1 and S3 for details).

To reach the quantum degenerate regime of polariton condensation, we use the sphere cap shaped indentations in the top DBR resulting in three-dimensionally confined, Laguerre-Gaussian type optical modes locally coupled to the perovskite sample. For excitation, we used 140 fs long laser pulses at 525 nm central wavelength and 80 MHz repetition rate. In Fig. 2a, we display a power-dependent series of PL spectra obtained with the lower polariton tuned to ~1.8 eV. We notice that this energy is slightly below the energy of the weakly coupled n=4 exciton (cf. Supplementary Section S2), which thus can contribute to the polariton population via intra-cavity pumping. As we increase the power by an overall factor of 20, we observe a pronounced rise in PL power by almost three orders of magnitude. From this data, we extract the input-output characteristic shown in Fig. 2b by fitting Lorentzian lines to the PL spectra. The non-linear increase in power displays the typical non-linear S-shape, observed in polariton lasing. A linear representation of the data is shown as an inset. From that plot, we extract a polariton lasing threshold of $P_{th} = 6.76\ fJ$ per pulse at a spot diameter of 1.44 µm. From the Lorentzian fits we also extract the central energy and emission linewidth plotted in Fig. 2c and d. We observe a blue shift of 0.36 meV extending across the threshold. Due to the femtosecond pulsed excitation, the expected collapse of the polariton linewidth is hard to discern [38], but we confirm the emergence of temporal coherence of the emission in the time-domain below (cf. Fig. 3). From the power density of the excitation at threshold, we estimate an injected density of electron-hole pairs of $\rho_{th} = 1.2 \cdot 10^{10}\ cm^{-2}$ (see Supplementary Section S4). The onset of polariton condensation thus occurs one order of magnitude below the Mott density of the n=3 2D HaP [10–13].

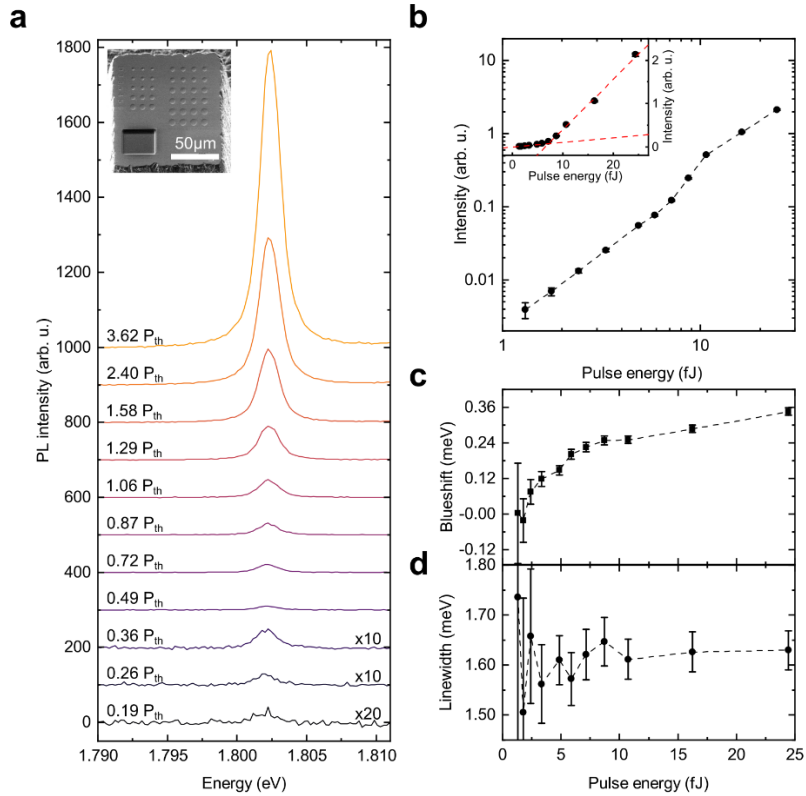

*Figure 2: **a** Power-dependent polariton emission spectra as a function of excitation pulse energy (525 nm central wavelength, 140 fs pulse duration, 80 MHz repetition rate). Spectra are vertically offset and multiplied by factors noted on the right for better visibility. Excitation powers are indicated on the left in multiples of lasing threshold. **Inset**: Electron micrograph of the upper cavity DBR shaped into a (100 µm)³ mesa to allow short cavity distances. Sphere cap-shaped indentations (6 µm diameter, 150 nm depth) result in three-dimensionally confined, discrete Laguerre-Gaussian type optical modes that we couple to the HaP flake. **b** Double logarithmic plot of the input-output curve extracted from the areas under Lorentzian fits to the spectra in a. **Inset**: Linear representation of the same data as in the main panel. From linear fits (dashed red) we extract a polariton lasing threshold of $P_{th} = 6.76\ fJ$ per pulse. (see Supplementary Section S5 for extended pulse energy range) **c** Spectral position of the Lorentzian fit as a function of pulse energy relative to the first data point. The emission undergoes a blue shift continuing beyond the lasing threshold. **d** Spectral linewidth extracted from the Lorentzian fits as a function of pulse energy.*

The blue shift of the lower polariton energy stems from several non-linear processes specific to the strong light-matter coupling regime. It is induced by the exciton-exciton (X-X) scattering due to Coulomb interactions and the non-linear saturation due to Pauli exclusion [39,40]. The X-X interaction constant can be estimated as $1.0\ \mu eV\ \mu m^2$ in a quantum well geometry, translating into the corresponding blue shift contribution for a given density and coupling (see Supplementary Section S6). The contribution of non-linear saturation can be estimated from the phase space filling, which scales as $\hbar \Omega A \sum_k |\phi(k)|^4 \approx 2.7\ \mu eV \mu m^2$, where $\hbar\Omega = 2g$, A and $\phi(k)$ are the Rabi splitting, the area of the sample and the exciton wavefunction in momentum space. Combining the two contributions, we estimate the non-linear interaction coefficient for the lower polaritons in the strongly red-detuned configuration to be $0.32\ \mu eV\ \mu m^2$. This agrees within an order of magnitude with the experimentally observed non-linearity deduced from Fig. 2c (see Supplementary Section S6 for details). Note that for densities above $\rho_{th}$, we observe a decrease in the slope of the lower polariton energy as the density increases, which is a typical feature of the onset of a polariton condensate.

To probe the emergence of long-range order and spatio-temporal coherence of the polariton mode [30,41–43], we image the emitted PL under strong optical pumping and quantify its spatial and temporal first-order coherence. This is achieved by inserting a Michelson interferometer into the detection path and placing a retroreflector at the end of one interferometer arm. This results in two spatially inverted, time-delayed copies of the emission overlapped at the exit of the interferometer. Adjusting the lateral offset of the retroreflector from the optical axis introduces an offset between the copy images in momentum space. When forming the PL image on a camera, this offset causes interference fringes from which we deduce the degree of spatial first-order coherence $g^{(1)}(\tau, r_\parallel)$ as a function of position $r_\parallel$ and interferometer delay $\tau$. Figure 3 panels a-c show the interference patterns for three powers below (panel a, $P = 0.9 P_{th}$), above (panel b, $P = 2.4 P_{th}$) and far above threshold (panel c, $P = 12.4 P_{th}$). Note that for better visibility the intensities in panels a and b were multiplied by factors of 5000 and 5, respectively. The dashed black circles mark the spatial extent of the sphere cap shaped mirror indentation, i.e. they give a rough estimate of the spatial extent for the uncoupled Laguerre Gaussian photonic mode that we couple to the sample. We observe that the coherent condensate substantially spreads outside the lens. By Fourier filtering, we extract the coherent portion of the signal from these images. The first order coherence is then obtained via the relation $g^{(1)}(\tau, r_\parallel) = \tilde{I}(\tau, r_\parallel)/2\sqrt{I_M(\tau, r_\parallel) I_R(r'_\parallel)}$ [30,44] with $\tilde{I}(\tau, r_\parallel)$ the Fourier-filtered interference pattern, $I_M(\tau, r_\parallel)$ the time-delayed image from the interferometer arm containing a planar mirror at its end and $I_R(r'_\parallel)$ the spatially inverted image from the other arm containing the retroreflector. The results of this calculation are shown in Fig. 3 d-f. We see a marked increase in the degree of spatial coherence between panels d and e, when increasing the excitation power above threshold. Above threshold, $g^{(1)}$ continues to increase slowly with power up to a maximum value of ~0.7 in panel f. To further quantify the coherence properties of the condensate, we extract $g^{(1)}$ above threshold from the maps in panels e and f and observe its evolution as a function of interferometer delay $\tau$. The results are plotted in Fig. 3 g-h. At $P = 2.4 P_{th}$ (panel g), the first order correlation function decays exponentially with a characteristic time scale of $\tau_c(2.4 P_{th}) = 0.71 ps$.

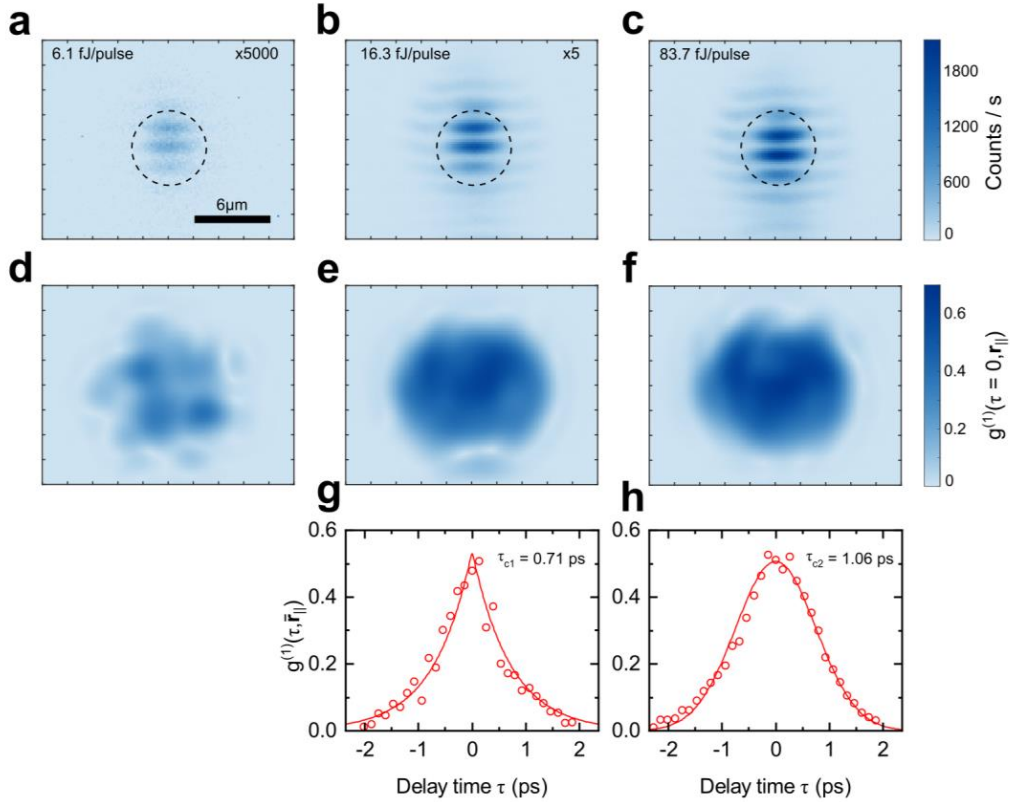

*Figure 3: Spatial coherence measurements of the emission from the fundamental Laguerre-Gaussian cavity mode for different excitation pulse energies: **a** below threshold (0.9 $P_{th}$), **b** above threshold (2.4 $P_{th}$), **c** far above threshold (12.4 $P_{th}$). Numbers in the upper right indicate multiplication factors for better visibility. **a-c** show spatially resolved interference images obtained by overlapping the emission from the mode with a spatially inverted, momentum-shifted copy realized with a Michelson interferometer using a retro-reflector in one arm. **d-f** show the spatial first-order correlation function at zero delay $g^{(1)}(\tau = 0, r_\parallel)$ extracted by Fourier analysis of panels a-c. **g(h)** shows the temporal first-order correlation function $g^{(1)}(\tau, \bar{r}_\parallel)$ averaged over $2\mu m \times 2\mu m$ at the center of the spatial coherence map in e(f) as a function of interferometer delay. The fits show coherence times of 0.71 ps (1.06 ps).*

Far above threshold at $P = 12.4 P_{th}$ (panel h), the shape of the decay notably changes and is well described by a Gaussian via $g^{(1)}(\tau, \bar{r}_\parallel) \propto \exp(\tau^2/\tau_c^2)$. This change is usually associated with number fluctuations in the condensate causing interaction-induced energy variations [36,45]. With this model, we obtain a 50% increase in coherence time far above $P_{th}$ resulting in $\tau_c(12.4 P_{th}) = 1.06 ps$. From cross-sections through the maps in Fig. 3 d-f, we extract the spatial extent of the first order coherence. The diameter is 7.8 μm below threshold, as extracted from panel d. It significantly increases to a diameter of 11.4 μm above threshold (1/e criterion), i.e. it becomes almost twice as big as the diameter of the sphere cap indentation.

## CONCLUSIONS AND OUTLOOK

Our study establishes 2D lead-based halide perovskites as a particularly well-suited material for non-linear polaritonics at room-temperature. We confirm polariton condensation at ambient conditions via the non-linear increase of quasi-particle population at exciton densities below the Mott density and the observed energy blue shift alongside with the onset of spatial and temporal coherence of the optical mode.

2D halide perovskites stand out in the landscape of polariton material platforms due to an unprecedented combination of key advantages: On the one hand, being chemically synthesized multi-quantum well stacks, they retain the unique flexibility in emission wavelength and oscillator strength of conventional QW stacks via controlling the thickness and number of inorganic layers, paired with excellent room-temperature operability. On the other hand, being 2D van-der-Waals materials, they share the benefits of easy hetero-integration typically found in TMDCs, opening an exciting new field of research for room-temperature polariton condensates based on hybrid 2D structures [21], integrated with electrically tunable cavities [46] or subjected to tunable photonic lattices [33,35,47]. In our study, hetero-integration has been instrumental to shield the 2D HaPs from photo-bleaching and moisture via hBN encapsulation. Based on the fast-paced progress with HaPs in the fields of photovoltaics and optoelectronics, we anticipate that the implementation of electrically driven room-temperature polariton lasers based on 2D HaPs is realistic in the near future resulting in vast opportunities for on-chip integration.

## ACKNOWLEDGEMENTS

The authors acknowledge support by the German research foundation (DFG) via the project SCHN1376 13.1. Financial support by the Niedersächsisches Ministerium für Wissenschaft und Kultur ("DyNano") is gratefully acknowledged.

O.K. and K.W.S. acknowledge the support from UK EPSRC Awards No. EP/X017222/1.

M.E. acknowledges funding by the University of Oldenburg through a Carl von Ossietzky Young Researchers' Fellowship.

H.P.A. acknowledges funding by the University of Valencia through the Margarita Salas grant (MS21-181) for the training of young doctors.

D.C. acknowledges the Ministry of Science and Culture (MWK) of the State of Lower Saxony and the Volkswagen Foundation ("Niedersächsisches Vorab - Research Cooperation Lower Saxony-Israel") and the Weizmann Institute of Science for support and and Dr. Sigalit Aharon (Weizmann Inst. and Princeton Univ.) for sample selection, logistics and guidance for their handling and preparation for measurements.

## METHODS

*Perovskite synthesis:*

Butylammonium, $C_4H_9NH_3$ (BA) methylammonium, $CH_3NH_3$ (MA) lead iodide $(BA)_2(MA)_2Pb_3I_{10}$ ($C_4N_3$) was crystallized using the slow-cooling method with minor modifications [7,9]. (see Supplementary Section S7 for detailed procedure)

*Sample preparation:*

DBR mirror coatings were prepared by HF sputtering of alternating layers of $SiO_2$ (107 nm) and $TiO_2$ (67 nm) terminating in $SiO_2$ with 10 mirror pairs at the bottom and 8 pairs at the top. Sphere cap shaped indentations in the top DBR (6µm diameter, 150 nm depth) were prepared by focused ion beam lithography (FIB, FEI Helios 600i) before mirror deposition.

(BA)$_2$(MA)$_2$Pb$_3$I$_{10}$ single crystals were micromechanically exfoliated with the scotch tape method followed by PDMS dry stamping. We first deposit a 10 nm thick hBN bottom layer (2D Semiconductors, substrate at 100°C, 5 min contact time), followed by the 2D HaP flake (substrate at 20°C, 20 min contact, PDMS GelPak grade zero, pre-baked at 85°C for 4h) and another 10 nm thick hBN flake to fully encapsulate the flake (substrate at 20°C, 20 min contact). The preparation was performed under yellow light and finished samples stored in the dark under N$_2$ atmosphere.

*Optical setup:*

Supplementary Figure S5 shows a schematic of the experimental setup. Femtosecond excitation pulses are derived from a Ti:Sapphire mode locked laser (Chameleon, Coherent) operated at 1050 nm and frequency-doubled in a BBO crystal (Crysmit optics, 21.3° cut, 0.1 mm thickness). Laser pulses are tightly focused onto the sample with a microscope objective (50X Mitutoyo Plan Apo NIR HR 0.65NA) through which we also collect the PL emission and measure WL reflectivity (Thorlabs SLS301). Details on the open cavity setup can be found in [32,47]. PL emission was analyzed with a Peltier-cooled CCD camera (Andor iXon Ultra 888) connected to a monochromator (Andor Shamrock 500i). Angle-resolved measurements were realized by imaging the back-focal plane of the objective onto the monochromator input with a 75X overall magnification. The Michelson interferometer consisted of a retroreflector (Thorlabs PS976M-B) in one arm and a planar mirror on a motorized stage (PI Linear Stage M-511) in the other. (see Supplementary Section S8 for further details)

# Supporting Information:

# Room-temperature polariton condensate in a two-dimensional hybrid perovskite

Marti Struve[1], Christoph Bennenhei[1], Hamid Pashaei Adl[1,2], Kok Wee Song[3], Hangyong Shan[1], Nadiya Mathukhno[1], Jens Drawer[1], Falk Eilenberger[4,5], Naga Pratibha Jasti[6,7], David Cahen[7], Oleksandr Kyriienko[3], Christian Schneider[1], Martin Esmann[1,+]

[1]Institut für Physik, Fakultät V, Carl von Ossietzky Universität Oldenburg, 26129 Oldenburg, Germany

[2]Instituto de Ciencia de los Materiales, University of Valencia, 46980 Valencia, Spain

[3]Department of Physics and Astronomy, University of Exeter, Exeter EX4 4QL, United Kingdom

[4]Fraunhofer-Institute for Applied Optics and Precision Engineering IOF, 07745 Jena, Germany

[5]Institute of Applied Physics, Abbe Center of Photonics, Friedrich Schiller University, 07745 Jena, Germany

[6]Department of Chemistry, Bar-Ilan Univ. Ramat Gan 5290002, Israel

[7]Department of Molecular Chemistry and Materials Science, Weizmann Institute of Science, Rehovot 7610001, Israel

[+]Corresponding author. Email: m.esmann@uni-oldenburg.de

## S1: Description of the coupled oscillator model

To model the observed polariton dispersion in Fig. 1c of the main text, we diagonalize a 5x5 coupled oscillator model based on the coupling Hamiltonian defined in Eq. (S1).

$$H = \begin{pmatrix} E_X & g & g & g & g \\ g & E_{cav,1} & 0 & 0 & 0 \\ g & 0 & E_{cav,2} & 0 & 0 \\ g & 0 & 0 & E_{cav,3} & 0 \\ g & 0 & 0 & 0 & E_{cav,4} \end{pmatrix} \quad (S1)$$

Here, $E_X = 2\,eV$ denotes the exciton energy of the n=3 layered halide perovskite (HaP), $g = 23.5\,meV$ is the light-matter coupling energy resulting in a normal-mode (Rabi) splitting of $\hbar\Omega = 2g$ in the simplest two-oscillator picture. To obtain accurate values for the uncoupled cavity dispersions $E_{cav,i}(d, \boldsymbol{k}_\parallel)$ as a function of cavity air gap $d$ and in-plane momentum $\boldsymbol{k}_\parallel$, we run transfer matrix simulations of the planar cavity system with the exciton resonance of the HaP switched off (cf. Supplementary Section S2). We find that a simple two-oscillator model is insufficient to describe our experimental observations due to the large normal mode splitting in comparison to the free spectral range of our microcavity and four photonic modes are required for the model to be accurate.

## S2: Angle-resolved photoluminescence

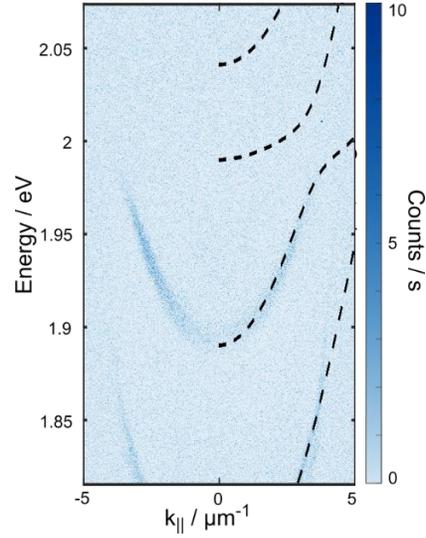

*Figure S1: Angle-resolved emission spectrum from planar portion of the DBR cavity shown in Fig. 1 of the main text with an air gap of 3.23 μm under CW excitation at 532 nm. The mode bounding 2 eV from below exhibits all features of a lower polariton branch as a result of an anti-crossing with the n=3 HaP exciton: An inversion point at $k_\parallel = 2.5~\mu m^{-1}$ and the associated characteristic reduction in group velocity towards higher in-plane momenta. A 5x5 coupled oscillator model (dashed lines) from which we deduce an exciton-photon coupling strength of $g = 23.5~meV$, accounts for the observed dispersion relation. No anti-crossing occurs with the exciton associated to the n=4 HaP at 1.9 eV. This exciton functions as an intra-cavity pump for the polariton experiments under strong optical driving.*

## S3: Transfer matrix simulations of white light reflectivity

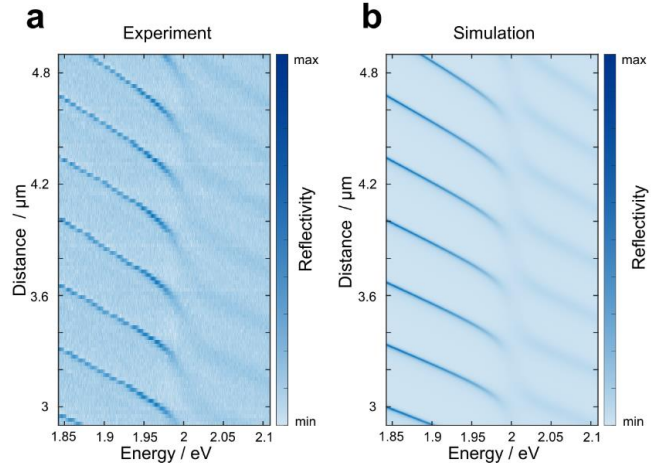

*Figure S2: **a** White light reflectivity (WL) spectra as a function of the cavity air gap at normal incidence ($k_\parallel = 0$). Same data as in Fig. 1c of the main text, but with extended distance range. **b** Corresponding transfer matrix simulation. We use this simulation to deduce accurate dispersions for the bare photonic modes in the coupled oscillator model by re-running the simulation with the exciton resonance in the dielectric function of the n=3 HaP switched off.*

## S4: Density of electron-hole pairs at the condensation threshold

At the polariton lasing threshold, the excitation conditions described in Fig. 2 of the main text (525 nm central wavelength, 140 fs pulse duration, 80 MHz repetition rate) result in an average number of $N_{ph} = 6.76\ fJ/2.36\ eV \approx 18{,}000$ photons reaching the sample surface per pulse.

By the same kind of transfer matrix simulations as shown Fig. S2, we determine that excitation light with in-plane momenta $|k_\parallel|/|k| < 0.3$ is efficiently coupled into the cavity through the first Bragg minimum of the DBRs while larger momenta are reflected. For the used microscope objective with NA=0.65, this results in a coupling efficiency of 20%. We furthermore find that the coupled portion is absorbed with 66% efficiency for the 218 nm thick HaP flake used in the experiments, this results in an average number of $N_X \approx 3{,}000$ excitons generated per pulse.

We estimate the number of optically active layers as follows: From the relative contributions to the PL spectrum in Fig. 1b of the main text, we estimate the number of layers with $n = 3$ and $n = 4$ in the sample to be the same (equal mixture of phases). From the transfer matrix simulation of our cavity, we deduce an optical penetration depth of $76\ nm$ for the pump wavelength into the perovskite sample. Assuming $n = 3$ layers and $n = 4$ layers are uniformly distributed across the penetrated depth with layer thicknesses of $2.6\ nm$ and $3.4\ nm$, respectively [1], we obtain $N_{HaP} \approx 13$ optically active layers. With a spot diameter of $d_{spot} = 0.82 \cdot 525\ nm/0.3 \approx 1.44\ \mu m$, this leads to an estimated exciton density of $\rho_{th} = 1.2 \cdot 10^{10} cm^{-2}$ at the threshold, i.e. one order of magnitude below the Mott density of the HaP material.

## S5: Input-output characteristic with extended range

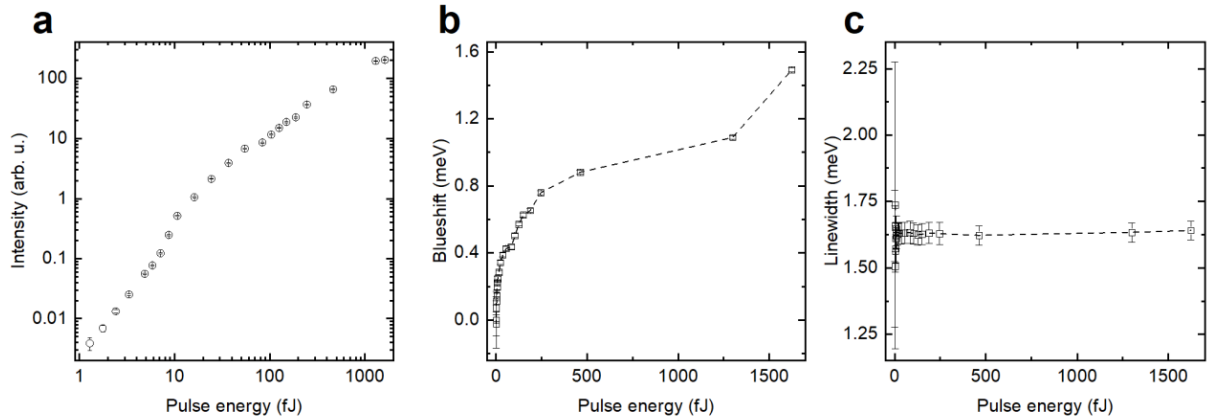

*Figure S3: **a** Double logarithmic plot of the input-output curve in Fig. 2b of the main text with pulse energy range extended up to $P = 240 P_{th}$. **b** Spectral position of the polariton emission as a function of pulse energy relative to the first data point. The emission undergoes a blue shift continuing beyond the lasing threshold with indications of an accelerated shift at very large pulse energies. **c** Spectral linewidth of the emission as a function of pulse energy.*

## S6: Theoretical model

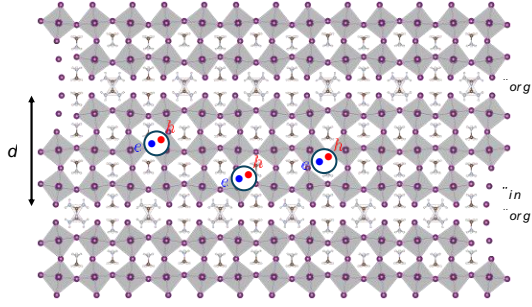

*Figure S4: Quantum well structure for theoretical modelling. The exciton bound states (white closed circles) are confined in the inorganic layers (middle region) with length $d$ and dielectric constant $\varepsilon_{in}$. The inorganic layer is encapsulated by the organic spacers (top and bottom regions) with dielectric constant $\varepsilon_{org}$.*

The properties of electrons and holes in 2D Ruddlesden-Popper perovskites can be approximated to a high precision as a confined motion in a quantum well of length $d$, see Fig. S4. Within this approximation, we can write the exciton wavefunction [2] as

$$X^+ = \int_{-\frac{d}{2}}^{\frac{d}{2}} dz\, dz' \sum_k \psi(k)\, \phi_e(z)\phi_h(z') a^+_{k,z} b_{k,z'},$$

where we separate its in-plane and out-of-plane variables. Here, $\psi(k)$ is the exciton wavefunction with $k$ being the relative in-plane momentum for the electron-hole pair, $\phi_{e/h}(z) = \sqrt{2/d}\cos(\pi z/d)$ is the lowest subband wavefunction for the electrons/holes, with $z$ being their out-of-plane coordinate. The in-plane part can be obtained by solving the Wannier equation [3]

$$\left[\frac{k_e^2}{2m_e} - \frac{k_h^2}{2m_h}\right]\psi(k) - \sum_q V(q)\psi(k+q) = E_b \psi(k),$$

where $m_e/m_0 = 0.097$ and $m_e/m_0 = 0.141$, with $m_0$ being the free electron mass [1], and the interacting kernel is

$$V(q) = \int_{-\frac{d}{2}}^{\frac{d}{2}}\int_{-\frac{d}{2}}^{\frac{d}{2}} dz\, dz'\, |\phi_e(z)|^2 W(q,z,z')\, |\phi_h(z)|^2.$$

The electron and hole interacting potential is obtained by solving the mesoscopic Poisson's equation [4–6], $\varepsilon(q^2 + \partial_z^2)W(q,z,z') = e/\varepsilon_0 \delta(z-z')$. The corresponding Coulomb potential is

$$W(q,z,z') = \frac{4\pi e^2}{\varepsilon_0}\frac{\cosh[q(\tfrac{1}{2}d-z)+\eta]\cosh[q(\tfrac{1}{2}d+z')+\eta]}{\sinh(qd+2\eta)}$$

with $\eta = \frac{1}{2}\ln\frac{\varepsilon_{in}+\varepsilon_{org}}{\varepsilon_{in}-\varepsilon_{org}}$. The dielectric constant for the inorganic layers and the organic spacer is $\varepsilon_{in} = 5.2$ and $\varepsilon_{org} = 2.2$ for $n = 3$ [7]. Using the Gaussian basis method [2], we obtain the binding energy $E_b = 180$ meV and the wavefunction $\psi(k)$.

In the cavity, the exciton and photon hybridize forming polaritons with the Hamiltonian

$$H_{pol} = \begin{bmatrix} \omega_c & \frac{1}{2}\Omega \\ \frac{1}{2}\Omega & E_X \end{bmatrix},$$

where $\omega_c$ is the cavity photon energy, $\Omega$ is the Rabi splitting, and $E_X$ is the exciton energy. This gives the polariton energies $E_\pm = \frac{1}{2}\left[\omega_c + E_X \pm \sqrt{(\omega_c - E_X)^2 + \Omega^2}\right]$.

The non-linear blueshift of the lower polariton energy $E_-(\rho_X)$ at increasing exciton density $\rho_X$ has two contributions. The first one originates from the Pauli blockade, as excitons are not pure bosons and consist of electrons and holes. The increased occupation reduces the available area (volume) to create excitons, leading to an effective reduction of the light-matter coupling. This in turn reduces the Rabi splitting as

$$\Omega(\rho) = \Omega_0 - \rho_X g_s.$$

The corresponding rate of reduction can be estimated from the excitonic wavefunctions calculated for our structure, leading to a rate of $g_s = A\,\Omega_0 \sum_k |\psi(k)|^4 \approx 2.7\,\mu eV\,\mu m^2$. The second contribution comes from the exciton-exciton Coulomb interactions. Here, the dominant contribution is the exchange between electrons and exchange between holes, leading to a shift of the exciton energy scaling in the lowest order as

$$E_X(\rho_X) = E_0 + \rho_X g_X,$$

where $g_X = 2\sum_{k,k'} V(k-k')[|\psi(k)|^2(|\psi(k')|^2 - \psi^*(k)\psi(k'))] \approx 1\,\mu eV\,\mu m^2$ [8]. Therefore, the blueshift of the lower polaritons at linear order in $\rho$ is

$$E_-(\rho) - E_-(0) \approx \frac{1}{2}\left[\left(1 + \frac{\omega_c - E_0}{\Delta}\right)g_X + \frac{\Omega_0}{\Delta}g_s\right]\rho_X$$

with $\Delta^2 = (\omega_c - E_0)^2 + \Omega_0^2$. From experiments, we have $\omega_c \approx 1.8\,eV$ for the cavity photonic mode and $E_0 \approx 2\,eV$ for the excitonic mode. Using the information above, we can estimate the non-linear interaction coefficient as

$$\frac{1}{2}\left[\left(1 + \frac{\omega_c - E_0}{\Delta}\right)g_X + \frac{\Omega_0}{\Delta}g_s\right] \approx 0.32\,\mu eV\,\mu m^2,$$

where we have effectively included the dependence on Hopfield coefficients for the strongly-detuned polaritonic mode. For the non-linear shift in Fig. 2c of the main text, we observe that the experimental estimate for the non-linear constant is $\delta_{th}/\rho_{th} \approx 1.5\,\mu eV\,\mu m^2$. This is within an order of magnitude to our theoretical estimate, which is on the conservative side. In fact, recent results from the diamagnetic coefficient measurements suggest that the Bohr radius of Ruddlesden Popper perovskite excitons is even larger than considered before (possibly due to screening and mass renormalization). This can explain the stronger interaction observed experimentally. Other points to consider include the need of independent density estimates and potential build-up of excitons in the reservoir. As the pump repetition rate is sufficiently high, we may induce a population of dark excitons that do not contribute to PL, but via phase space filling influence the non-linear shifts. We consider that further research in this direction, as well as treatment of disorder physics can help elucidating the nature of polaritonic non-linearity in 2D Ruddlesden Popper perovskite condensates.

## S7: Perovskite synthesis

The crystals were grown as follows below and characterized as described in Refs. [9,10]. In short, butylammonium, $C_4H_9NH_3$ (BA) methylammonium, $CH_3NH_3$ (MA) lead iodide ($BA_2MA_2Pb_3I_{10}$, $C_4N_3$) was crystallized using the slow-cooling method with minor modifications [1]. 5.045 mmol (1.126 g) PbO (ACS reagent, 99.0%, Sigma-Aldrich) was dissolved in 5mL HI (57% in $H_2O$, Sigma-Aldrich) and 850 µL hypophosphorous acid solution (50 wt. % in $H_2O$, Sigma-Aldrich) in an 18 mL vial. After tightly screwing the vial's cap, the mixture was stirred (magnetic stirrer) and heated on a hot plate that was set to 110 °C. The color of the mixture changed from black to clear yellow within a minute. The stirring and heating continued until full dissolution of the PbO (1-2 hours). In the meantime, in an ice-bath, 3 mL of cooled HI (at 4 °C) were mixed with 98 µL butylamine (99.5%, Sigma-Aldrich) by vigorously stirring the HI with a magnetic stirrer and adding the butylamine dropwise. 0.477 g methylammonium iodide (MAI, GreatCellSolar materials), were added to the HI prior to the butylamine addition. The vial was then tightly sealed and stirring continued until no vapor was seen in the upper part of the vial. Once the two mixture was ready, the HI+butylamine+MAI mixture was added, dropwise, to the Pb-containing vial while continuing to stir and heat. This led to the formation of a dark red powder in the vial. Then the vial was tightly sealed and heated, while stirring, until the dark red powder fully dissolved (the hotplate was set for this purpose to 140-170 °C for 10-20 minutes). Once the solution was perfectly clear, we carefully took the magnet out of the vial, tightly sealed it again, and transferred it into a system for controlled slow cooling that contained a silicone-oil bath, which was preheated to 105 °C. The silicone oil bath sat in a closed (Pyrex) glass container to maintain a uniform temperature. The temperature of the silicone oil bath was gradually decreased to RT at a rate of 1°C /h. Once the cooling process was completed, large black plates of $BA_2MA_2Pb_3I_{10}$ were seen in the bottom of the vial. The single crystals were then taken out by evacuating the supernatant and drying them gently and thoroughly with fiber free blotting paper. The crystals were then evacuated overnight in the anti-chamber of a glove box and stored in a N2-filled glovebox. Structural, optical and electronic characterizations of the crystals used here can be found in Refs. [9,10].

## S8: Optical setup

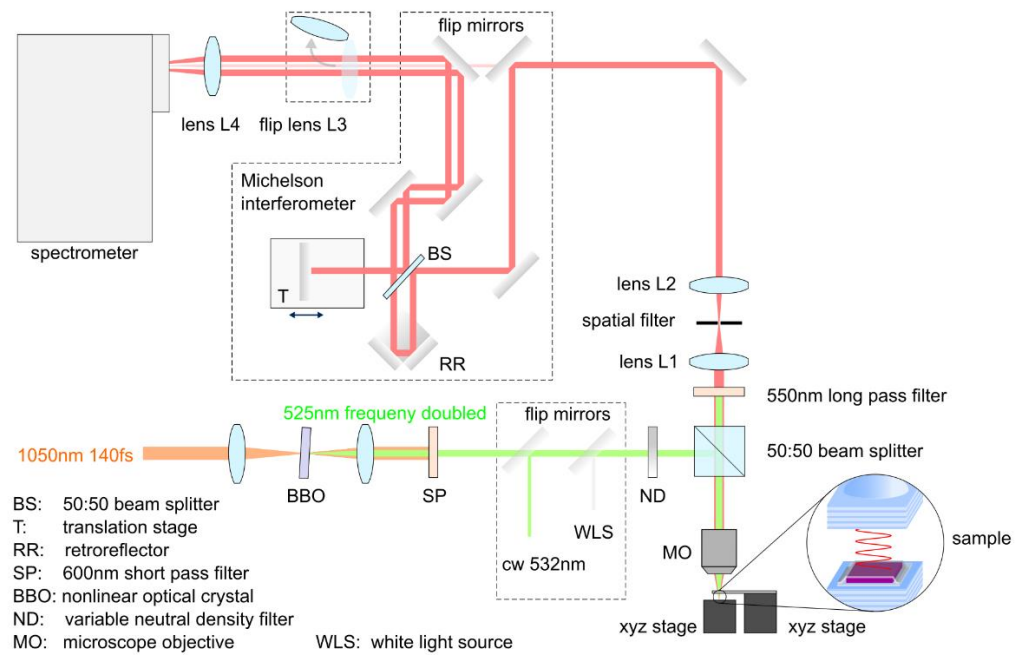

*Figure S5: Schematic of the experimental setup used to measure the spatial coherence of the polariton condensate.*

**Excitation path**: Femtosecond pulses from a mode locked Ti:Sapphire laser source (Chameleon, Coherent) operated at 1050 nm with 80 MHz repetition rate, are focused on a non-linear optical crystal (BBO, Crysmit optics, 21.3° cut, 0.1 mm thickness) to generate frequency doubled pulses at 525 nm. The light from the BBO is collected and collimated by a lens, before a 600 nm short pass filter isolates the frequency doubled light. For white light reflection or linear photoluminescence measurements, light from a white light source (Thorlabs SLS301) or from a continuous wave DPSS laser (532 nm) can be coupled in the excitation path by flip mirrors (dashed box). The excitation power can be controlled with a motorized variable neutral density filter. The excitation light is reflected by a 50:50 beam splitter towards a microscope objective (50X Mitutoyo Plan Apo NIR HR 0.65NA), which focuses the light on the sample. For polariton experiments, the active material is prepared in an open cavity configuration as described in *[11]*. Dispersion by the beam splitter and the microscope objective lead to a pulse length of the frequency doubled pulses of about 140 fs.

**Collection path**: Operating in reflection geometry, the same microscope objective collects the emission from the sample (microcavity) and transmits it through the beam splitter. Reflected excitation laser light is filtered out by a 550 nm long pass filter. Lens L1 with its focal point positioned on the back aperture of the microscope objective focuses the emission to a spatial filter, where lens L2 in confocal configuration collimates the light again. Lens L4 images the emission onto the entrance slit of an imaging spectrometer (Andor Shamrock 500i), with a Peltier-cooled EMCCD camera (Andor iXon Ultra 888, operated without EM gain). Adding flip lens L3 (top left dashed box) in a Fourier imaging configuration allows for angle-resolved measurements, by imaging the back-focal plane of the microscope objective onto the entrance slit of the spectrometer.

**Michelson interferometer**: Additionally, a Michelson interferometer can be added to the collection bath by two flip mirrors (large dashed box). The beam is split by a 50:50 beam splitter (BSW10R), with the reflected part again being reflected back to the beam splitter by a retroreflector (PS976M-B) generating a spatial offset to the transmitted beam, that is reflected back to the beam splitter by a planar mirror on a motorized translation stage (PI Linear Stage M-511). The retroreflector introduces a point reflection with respect to the beam reflected by the planar mirror. Varying the position of the translation stage adds a temporal offset between the beams. Furthermore, the spatial offset between the two beams creates a difference in wave vectors between the two beams when focused onto the spectrometer slit by lens L4. This creates a time delay-dependent fringe pattern on the CCD camera, which is used to extract information about the spatial coherence.